# Foams with 3D Spatially Programmed Mechanics Enabled by Autonomous Active Learning on Viscous Thread Printing

*Brett Emery†[1] Orcid: 0000-0002-3163-225X*

*Kelsey L. Snapp† Orcid: 0000-0001-5984-0723*

*Daniel Revier Orcid: 0000-0001-6246-7819*

*Vivek Sarkar Orcid: 0009-0000-4120-8973*

*Masa Nakura Orcid: 0009-0000-6210-1702*

*Keith A. Brown Orcid: 0000-0002-2379-2018*

*Jeffrey Ian Lipton* Orcid:0000-0003-0843-0999*

*Affiliations*

B. A. Emery, J. I. Lipton*

Department of Mechanical and Industrial Engineering, Northeastern University, 815 Columbus Ave, Boston, MA 02120, United States

E-mail: emery.b@northeastern.edu, j.lipton@northeastern.edu

K. L. Snapp, K. A. Brown

Department of Mechanical Engineering, Boston University, 110 Cummington Mall College of Engineering, Boston, MA 02215, United States

E-mail: ksnapp@bu.edu, brownka@bu.edu

D. Revier, V. Sarkar, M. Nakura

Department of Computer Science and Engineering, University of Washington, 185 E Stevens Way NE, Seattle, WA 98195, United States

E-mail: drevier@cs.washington.edu, viveksar@uw.edu, mnakura@cs.washington.edu

**Data Availability Statement:**

All data will be available through the Northeastern University data repository service (DRS)

---

[1] † Co-first authors

**Funding Statement:**

This work was funded by: a gift from the Ford Motor Company, NSF Grant number 2212049, the National Defense Science and Engineering Graduate Fellowship (NDSEG) program. KAB and KLS acknowledge support from the National Science Foundation (DMR-2323728) and the Rafik B. Hariri Institute for Computing and Computational Science and Engineering.

**Conflict of Interest Disclosure:**

The authors have a patent application for path planning methods used (Patent Application 63/527,296 filed 7/17/2023)

Abstract:

Foams are versatile by nature and ubiquitous in a wide range of applications, including padding, insulation, and acoustic dampening. Previous work established that foams 3D printed via Viscous Thread Printing (VTP) can in principle combine the flexibility of 3D printing with the mechanical properties of conventional foams. However, the generality of prior work is limited due to the lack of predictable process-property relationships. In this work, we utilize a self-driving lab that combines automated experimentation with machine learning to identify a processing subspace in which dimensionally consistent materials are produced using VTP with spatially programmable mechanical properties. In carrying out this process, we discover an underlying self-stabilizing characteristic of VTP layer thickness, an important feature for its extension to new materials and systems. Several complex exemplars are constructed to illustrate the newly enabled capabilities of foams produced via VTP, including 1D gradient rectangular slabs, 2D localized stiffness zones on an insole orthotic and living hinges, and programmed 3D deformation via a cable driven humanoid hand. Predictive mapping models are developed and validated for both thermoplastic polyurethane (TPU) and polylactic acid (PLA) filaments, suggesting the ability to train a model for any material suitable for material extrusion (ME) 3D printing.

## 1. Introduction

Foams are ubiquitous for a wide range of common applications ranging from upholstery, packaging, and personal protective equipment, as well as more exotic functions such as medical implants, vibration attenuation, and structural lightweighting[1–3]. Many of these applications primarily use homogeneous foams due to the prevalence of chemical manufacturing techniques used to produce large blocks of material[2–10]. When a single uniform foam is unable to meet design requirements the conventional approach is to rely on adhesive based lamination. This process is consequently restricted to creating discrete boundaries between property zones, resulting in composites with stress concentrations and discontinuities. Spatially resolved manufacturing processes, such as 3D printing, are not limited in this way.

Material extrusion (ME) is a common form of 3D printing in which a material, such as a thermoplastic, is deposited layer by layer according to a prescribed geometry in order to form a volumetric part. A key feature of ME is the ability to spatially vary mechanical properties by patterning the underlying microstructure of an object. This is used for applications such as acoustic/elastic cloaking, patterning metamaterial mechanisms, personalizing biomechanical devices, etc.[11–15]. This ability to locally control material properties offers a unique approach to improving the functionality and manufacturability of foams. However, previous ME processes have struggled to manufacture foams due to various fundamental challenges.

ME produced foams have previously relied on three methods to produce foams: 1) using secondary foaming materials, 2) directly injecting gas to produce bubbles, or 3) explicitly programming cellular geometries. Each results in a limited range of pore sizes and stiffnesses which then restricts the ability to spatially vary properties[7,16–19], as well as other process specific limitations. Foaming materials are constrained by the composition of chemical gassing/blowing agents or physical porogen inclusions (i.e. Syntactic Foaming[19]), limiting material compatibility and pore size control. This inflexibility in varying properties across foams makes it primarily suitable for producing large foam monoliths or hybridizing with other foam production processes[7,17,19,20]. Bubble-based ME processes such as direct foam writing[10] and direct bubble writing[16] enable individual cell characteristics to be tailored locally while maximizing material usage efficiency. However, bubble-based processes remain limited in material selection and lack fine detail resolution. Finally, for explicit cell geometry, the printer's resolution must be finer than the cellular structure's unit cell, typically by an order of magnitude or more[18,21,22]. This design constraint increases fabrication time, introduces the need for support material, and limits applicability.

In comparison, viscous thread printing (VTP) leverages the physical phenomenon of viscous thread instability to produce open cellular materials without explicit cell design or complex pathing algorithms[23]. VTP uses the dynamics of the printing process to control the microstructure of the object and consequently the underlying mechanical properties[23–25]. Existing work using similar and precursor methods to VTP have established its viability with various materials for soft robotic gripping[26], pressure sensing and proprioception[27], lightweight circuitry[28], food printing[29], architecture[30,31], and artistic expression[32,33]. Our previous work with VTP created open cellular foams with homogenous and limited gradient Young's moduli using TPU on standard ME printers, and demonstrated a reduction in print artifacts and boundary defects, as well as enhanced failure performance over the previous state of the art [23,24]. However, establishing the correlation between VTP inputs and resultant material properties has required iterative manual testing of each parameter combination to determine their respective material properties and did not result in generalizable or transferrable models. By producing foams via VTP, we avoid the need for intricate design files or extended fabrication times, preserve the full selection of ME-compatible materials, expand the ability to locally program mechanical properties, and offer an extensive cell size range.

Here, we develop a predictive model that links input VTP parameters to output foam material properties. Through this model, we discover and establish a parametric subspace which produces homogeneous foams by accurately predicting structural properties like effective layer height with $R^2 = 0.980$. We realized a Gaussian process regression (GPR) model able to accurately map VTP coiling parameters to resulting material stiffness by exploring and characterizing the available property space through a data-driven approach. To do this, we employed a self-driving lab (SDL), which uses automation and active learning to perform experiments in high dimensional space without human intervention [34,35]. This model was then validated using leave one out cross validation (LOOCV) to predict the effective modulus with an $R^2$ value of 0.946. An additional GPR model, based on input VTP parameters, was developed from principal component analysis (PCA) of the full stress-strain curves of foams to more broadly predict their mechanical properties. These models allow the rapid development of homogenous foams with targeted material properties. By applying these models continuously, they additionally facilitate gradient foams where the properties intentionally and predictably vary in space. Several complex exemplars were constructed to illustrate the newly enabled capabilities of foams produced via VTP, including 1D discrete and linear gradients on rectangular slabs, 2D localized stiffness zones on an insole orthotic and living hinges, and programmed 3D deformation via a cable driven anthropomorphic hand.

## 2. Result and Discussion

### 2.1. VTP Process Description

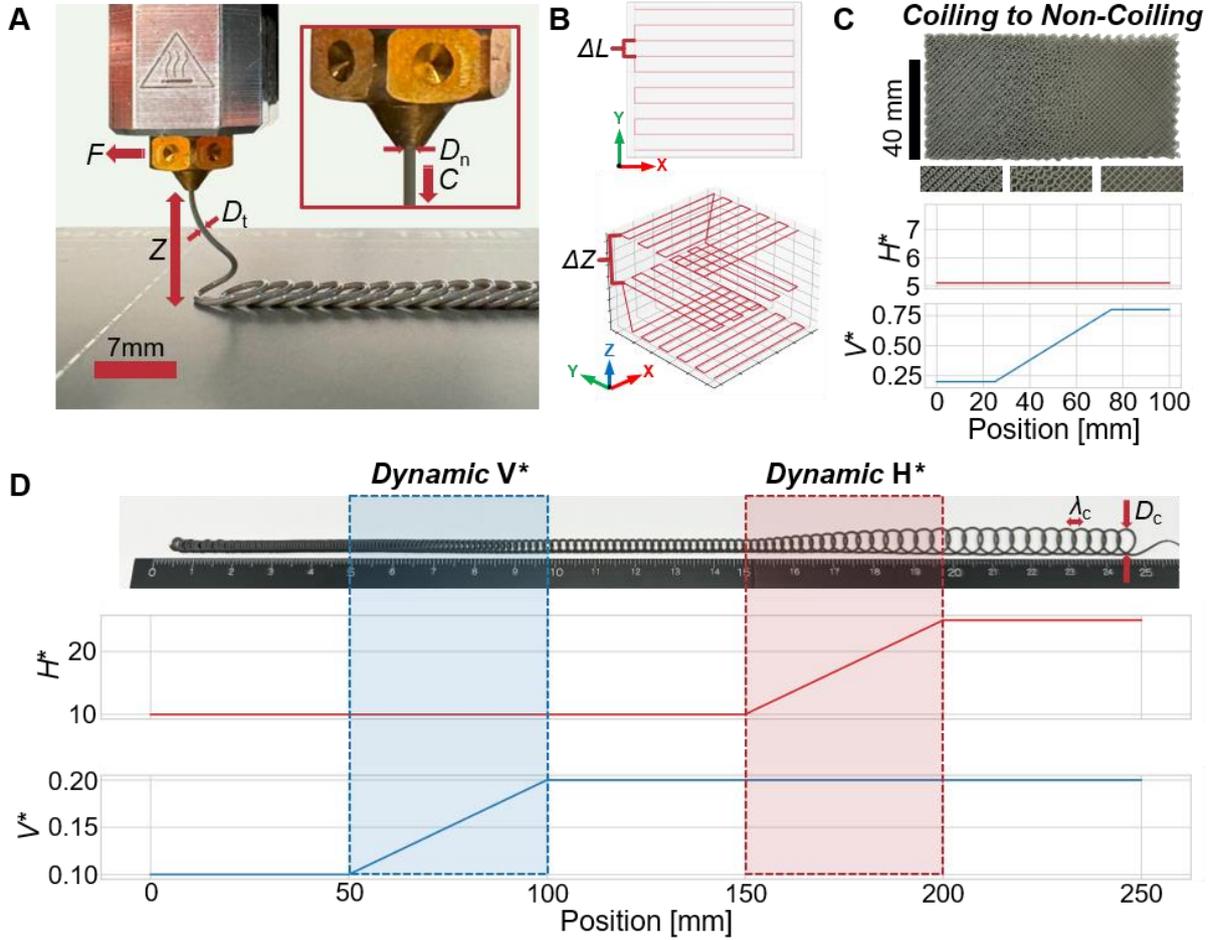

**Figure 1.** A) Typical material extrusion (ME) 3D printing setup during viscous thread printing (VTP), annotated with relevant parameters: nozzle translation speed ($F$), nozzle height ($Z$), thread diameter ($D_t$), nozzle diameter ($D_n$), and thread extrusion speed ($C$). B) 2D plot and 3D plot of a typical single-layer and multi-layer rectilinear toolpath, with annotated line spacing ($\Delta L$) and layer spacing ($\Delta Z$) respectively. C) Images of coiling behavior transitioning from coiling to non-coiling ME, with close-up images as seen on the left, middle, and right portions of the specimen, and corresponding plots of $V^*$ and $H^*$ as a function of position in x [mm]. D) 1D line of translating coiling with dynamic coiling behavior changes induced via $V^*$ and $H^*$ manipulation, annotated with coil wavelength ($\lambda_c$) and coil diameter ($D_c$), and corresponding $H^*$ vs. Position x [mm] and $V^*$ vs. x [mm] plots with respect to coiling behavior seen in C.

Fundamentally, VTP exploits the tendency of viscous liquids in the form of slender threads to coil upon deposition. This phenomenon, known as viscous thread instability, has been thoroughly characterized in 1D and 2D[36–40] and is exemplified by the coiling pattern formed when honey is poured onto a surface. By applying this principle on conventional ME machines, a stationary nozzle produces consistent coiling behavior at scales smaller than the printer's native resolution. Furthermore, by following a simple linear toolpath, a continuous line of coils may be produced (Figure 1A) which enables smooth and continuously varying material properties in space. This toolpath may be arranged in a simple rectilinear pattern in order to produce 2D meshes, which are then successively stacked to create three-dimensional structures as seen in Figure 1B. The resulting open-cellular microstructure's geometry is stochastically distributed yet consistent with respect to its constituent coiling behavior, provided the printing conditions are maintained.

There are two key parameters which characterize VTP coiling behavior in 1D as described by [37,39,41,42]: the dimensionless velocity $V^*$ and the dimensionless height $H^*$, defined as

$$V^* = \frac{F}{C} \tag{1}$$

$$H^* = \frac{H}{D_t} \tag{2}$$

where $F$ is the translation speed of the printhead, $C$ is the exit speed of the material from the nozzle, $H$ is the height of the nozzle above the substrate, and $D_t$ is thread diameter. The $D_t$ is often larger than the nozzle diameter $D_n$ due to die swell, and can be calculated using the die-swell parameter $\alpha$ using the equation $D_t = \alpha D_n$. Each of these variables are controlled either directly or indirectly through commands provided to the printer via G-code, a numerical control programming language that provides precise instructions for the movements, speeds, and coordination of the print head and extruder during fabrication. Of the parameters which constitute $V^*$ and $H^*$, only $F$ and $Z$ are explicitly provided in G-code. $C$ may be calculated as a function of the linear filament feed rate due to the assumption that the filament volume flow into the nozzle is equal to the thread volume flow out of the nozzle[23], and $\alpha$ is a material- and process-dependent constant that is determined empirically.

Maintaining control of coiling geometry is essential for the purposes of realizing printable foams. Prior work has shown that $V^*$ is inversely proportional to linear coil density and therefore has a first order effect on coiling wavelength $\lambda_c$, while $H^*$ is directly proportional to coil diameter $D_c$ and has a second order effect on $\lambda_c$ [37,39,41]. Coiling mode refers to the many types of distinct coiling patterns which may be produced by manipulating $V^*$ and $H^*$ such as translating coiling, alternating coiling, and equidimensional [37,39,41]. Of these modes, translating coiling and alternating coiling represent the majority of useful outcomes, as they are the only two modes that produce consistent, self-contained coils, which form the basis of the cellular structure characteristic of foams. However, the equidimensional mode also warrants consideration. Despite its lack of features resembling cellular structures, it serves as a bridge between VTP coiling and conventional ME non-coiling processes as seen in

Figure 1C. This is because both the equidimensional mode and conventional ME deposition occupy only the space directly beneath the nozzle along the prescribed toolpath. As such, the coiling mode parameters $V^*$ and $H^*$ have the most influence over microstructure which, beyond base material properties, is the primary driving factor behind the foams' material properties.

By varying $V^*$ and $H^*$ as a function of space, we may continuously vary coiling mode and therefore locally control microstructure (Figure 1C, 1D). In order to practically use the aforementioned coiling modes to produce a coherent volume, the extruding nozzle must follow a specified toolpath. For maximum generality, we have chosen to maintain a simple rectilinear toolpath with each consecutive layer being rotated 90º (Figure 1B). The shape of this rectilinear toolpath is determined by the variables $\varDelta L$ and $\varDelta Z$, referring to the spacing of parallel lines and Z-height increment of sequential layers, respectively. Both of these parameters have significant influence on the overall properties of the resulting part as they dictate where material is deposited. However, due to self-interaction between deposited threads and preceding layers, predicting the exact geometry of the deposited thread and its aggregate structures is challenging. Consequently, determining the impact on the overall foam structure resulting from tuning any or all VTP parameters has traditionally necessitated explicit testing between iterations.

## 2.2. Modeling and Homogeneous Subspace

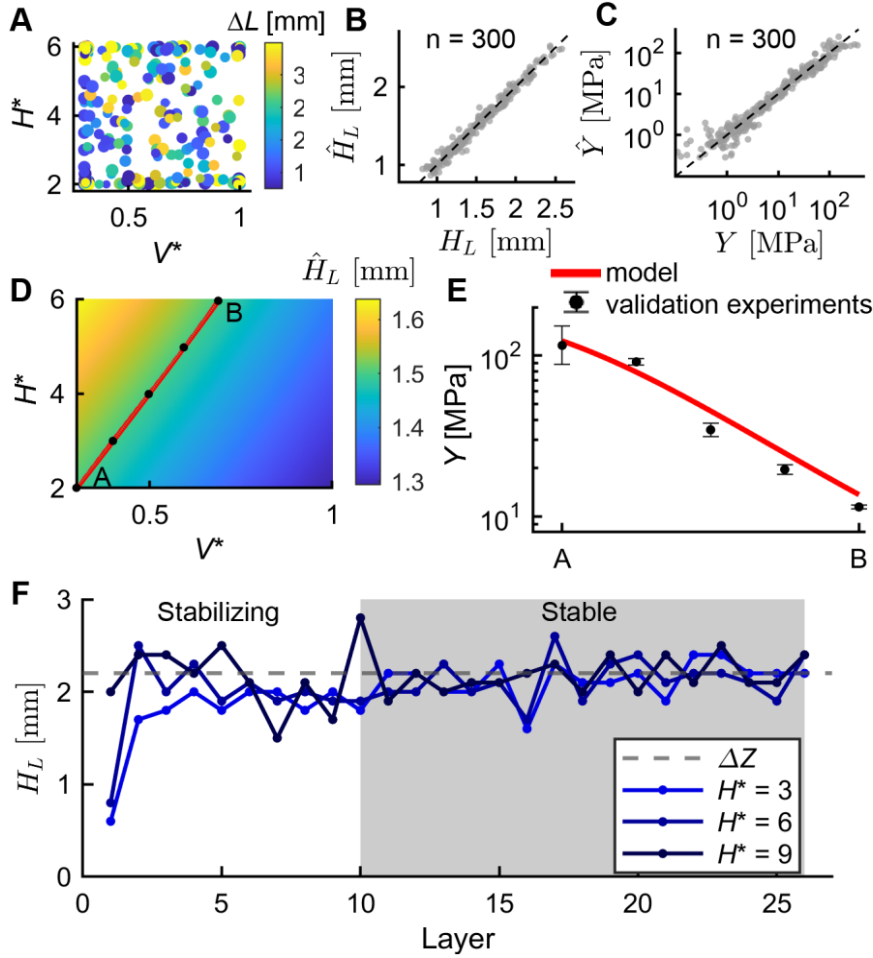

**Figure 2.** A) $H^*$ vs. $V^*$ for all 300 experiments performed in PLA with color indicating $\varDelta L$ and marker size indicating $\varDelta Z$. B) Parity plot of predicted layer height $\hat{H}_L$ vs. layer height $H_L$ for Gaussian process regression (GPR) model using leave one out cross validation (LOOCV). C) Log-log parity plot for predicted effective modulus $\hat{Y}$ vs. effective modulus $Y$ for the GPR model using LOOCV. D) Slice in 2D of $H^*$ and $V^*$ where $\varDelta L$ and $\varDelta Z$ are equal to 1.5 mm. Color indicates $\hat{H}_L$ and the red line indicates the predicted homogenous subspace where $\hat{H}_L = \varDelta Z$. Black dots indicate five equally spaced samples selected for validation testing. E) GPR model's $\hat{Y}$ (red line) vs. results of five validation experiments (black). Error bars represent one standard deviation in semi-log space from each condition being tested in triplicate. F) $H_L$ vs. layer number for three PLA cubes with different $H^*$. After several layers, $H_L$ for each cube stabilizes to 2.2 mm, which is equal to $\varDelta Z$.

While the flexibility of VTP is a virtue from a materials design perspective, tuning the processing parameters to realize homogenous materials presents a challenge. Specifically, when the first layer of a sample is printed, the processing variables $V^*$, $H^*$, and $\varDelta L$ collectively determine the layer height $H_L$. For a VTP structure to be homogeneous, $\varDelta Z$ must equal $H_L$ so that the distance between the print nozzle and substrate does not change when printing the next layer, ensuring the same average microstructure from layer to layer. Therefore, for each $V^*$, $H^*$, and $\varDelta L$ there is a unique $\varDelta Z$ that will lead to a homogeneous foam. The subspace in which these conditions are met is referred to as the homogeneous subspace.

In order to predict the correct $\Delta Z$ for each $V^*$, $H^*$, and $\Delta L$ triplet, we used an SDL to run an experimental campaign that iteratively selected and performed 300 experiments using Bayesian optimization to minimize the uncertainty across the full four-dimensional space on VTP foams formed out of polylactic acid (PLA) (Figure 2A). After these experiments, the performance of the final Gaussian process regression (GPR) model to predict $H_L$ was evaluated using leave one out cross validation (LOOCV) and found to have a $R^2$ value of 0.980 (Figure 2B). In addition, we also conditioned a GPR to predict an effective modulus $Y$ of each part, which was found to have a LOOCV $R^2$ value of 0.946 (Figure 2C). Importantly, these two models can be combined to allow precise control of the properties of a homogenous cube by enabling inverse design to target a desired $Y$.

We hypothesized that the SDL-derived models for $\hat{H}_L$ and $\hat{Y}$ could allow predictive control over the VTP outcome and enable us to fully exploit the capabilities of VTP printing. In particular, a major goal is to spatially vary the mechanical properties of a material in a manner that is easy and reliable to fabricate. One approachable way to implement this would be to maintain a constant $\Delta L$ and $\Delta Z$ throughout the part, thus enabling a consistent rectilinear toolpath as shown in Figure 1B, while maintaining the utility of $V^*$ and $H^*$ to change the $Y$ of different areas of the part. To explore this possibility, a 2D slice of the $H^*$ and $V^*$ parameter space at a constant $\Delta L$ and $\Delta Z$ can be visualized (Figure 2D). The $\hat{H}_L$ is shown, with the red line indicating the homogenous subspace for which $\hat{H}_L = \Delta Z$. Note that while this red line appears to be linear, the underlying GPR model makes no assumption of linearity. Five points that are equally spaced from A to B are selected along this homogenous line. The $Y$ of the physical experiments is then compared to the predicted effective modulus $\hat{Y}$ from the GPR model, showing that the it can be modulated between ~10 and 100 MPa while remaining in the homogenous subspace with constant $\Delta L$ and $\Delta Z$ (Figure 2E).

While the presence of a smooth homogenous subspace that facilitates the manufacture of graded materials is a powerful outcome of this study, the shape of the inhomogeneous region also revealed a fortuitous relationship in the underlying VTP process. Specifically, Figure 2D shows that there is a positive correlation between $H^*$ and $H_L$. This positive correlation implies the existence of a stabilizing basin of attraction to correct errors in the selection of $\Delta Z$. If $\Delta Z$ is too large, then $H_L < \Delta Z$ and the effective $H^*$ will increase each layer until $H_L = \Delta Z$. In contrast, if $\Delta Z$ is too small, then $H_L > \Delta Z$ and the effective $H^*$ will decrease until $H_L = \Delta Z$. Consequently, the component will homogenize at a new effective $H^*$ if the selected parameters are not already in the homogenous subspace, although this process will change the properties of the foam and produce a non-homogenous zone for several layers during stabilization. For this reason, it is important to determine the correct *ΔZ* before printing. Nevertheless, the basin of attraction means that small errors in the selection of *ΔZ* will not be catastrophic to the printing process and that any anomalies during printing will self correct. To demonstrate this basin of attraction, we printed three samples with the same *V\**, *ΔL*, and *ΔZ* with variations in their *H\**. After printing these samples, we measured the thickness of each layer. As expected, $H_L$ for each of the experiments stabilized at the selected *ΔZ*, regardless of *H\** (Figure 2F).

## 2.3. Dynamic Coiling and Gradient

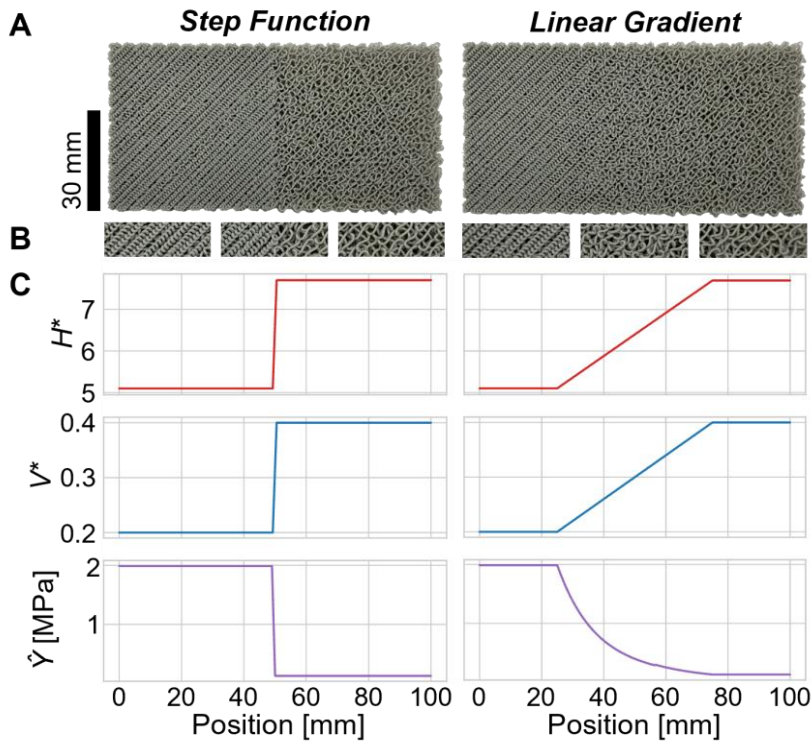

**Figure 3.** A) Rectangular specimens with constant layer heights, as modeled by the homogeneous subspace, demonstrating gradient coiling behavior. The specimens transition from high to low density from left to right, exhibiting both discrete and linear changes in coiling behavior respectively. B) Close-up images of coiling behavior observed in the left, middle, and right portion of each specimen, respectively. C) Corresponding plots of *H\**, *V\**, and *Ŷ*, from top to bottom, as a function of position in *x* [mm].

Having isolated the homogeneous subspace, we can query it to maximize the available property space of compatible homogeneous foams. While both *V\** and *H\** affect microstructure density, $V*$ is the primary factor in determining linear coil density and, consequently, has the strongest impact on overall VTP foam density. As such, previous work was able to establish that continuous gradients could be achieved by singularly varying $V*$[23]. However, without also varying $H*$ to maintain layer thickness homogeneity, the range of mechanical properties that can be achieved in these foams, without significant structural consequences, is severely limited. Only by applying the homogeneous subspace model pioneered by this work, is the full VTP mechanical property space able to be navigated without inducing unintentional layer thickness variations. Furthermore, the application of this model enables the ability to guarantee layer thickness compatibility between distinct microstructures. Thereby allowing for the programming of zones with specific desired microstructure dependent mechanical properties, and the creation of bespoke transition regions such as discrete boundaries or continuous gradients as seen in Figure 3, without unintended structural consequences.

## 2.4. Exploration of Enabled Capabilities

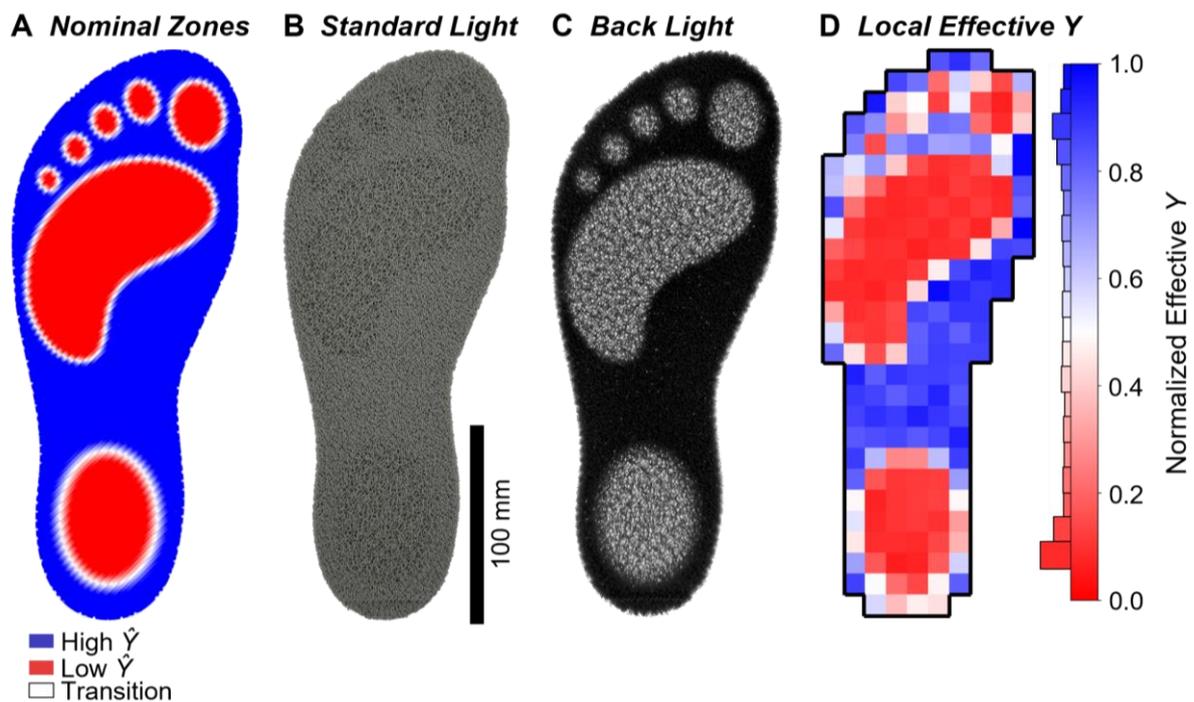

**Figure 4.** A) Density color coded toolpath of example multi-stiffness VTP orthotic. B) Top view of printed orthotic under standard lighting conditions. C) Top view of printed orthotic when illuminated from beneath. D) Normalized local effective $Y$ map of orthotic with respective colorbar and occurrence histogram.

By applying gradients to induce specific output material properties, a wide variety of potential application spaces emerge. For example, fine control over compressive stiffness is vital to applications such as padding, cushioning, or pressure relieving apparatus. To explore whether VTP as a method to tailor stiffness could be used in these contexts, we designed and printed the custom orthotic seen in Figure 4. We validated the stiffness for this orthotic by conducting localized compression tests. This generated the local effective *Y* map seen in Figure 4D. We normalized the modulus to highlight relative extremes in modulus. Additionally, due to the relationship between microstructure, density, porosity, and stiffness in cellular materials [2] for relatively thin objects, such as the orthotic in Figure 4, the effect of VTP programmed microstructure zones also manifest in the form of different surface textures as seen in Figure 4B, as well as transparency due to porosity in Figure 4C.

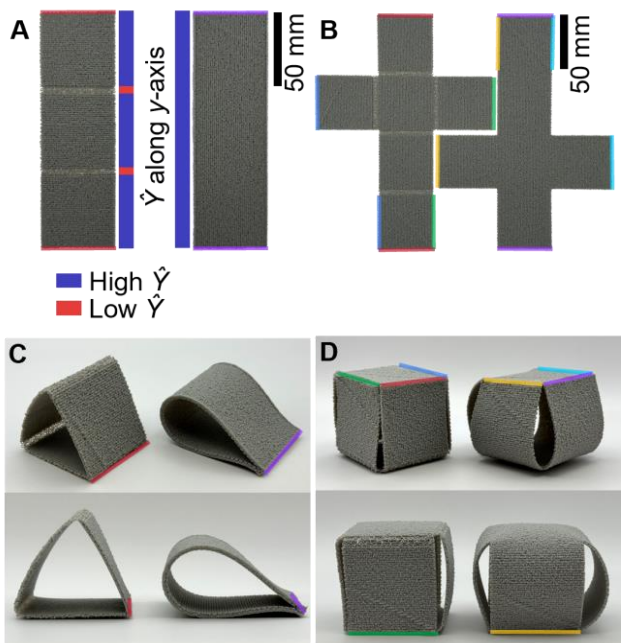

**Figure 5.** Examples of folding nets where the left column of each quadrant contains specimens with VTP compliant hinges, while the right are the same nets without hinges. Fixed edges are color coded respectively. A) Top view of unfolded triangular nets with annotated $\hat{Y}$ zones along *y*-axis. B) Top view of unfolded box nets. C) Isometric and front views of folded triangular nets. D) Isometric and front views of folded box nets.

Programmed deformation is an additional capability enabled by programmed density and stiffness which we explore in both 2.5D and 3D objects. In existing literature programmed deformation in 2D, 2.5D, and 3D have been applied for use in examples such as metamaterial doorknob mechanisms and pliers [14,43], as well as for use in soft robotics [27]. Utilizing VTP's ability for programmed stiffness, we show that the local reduction of stiffness enables selective motion through elastic deformation in the form of compliant joints and hinges [44,45] as seen in Figure 5. Due to the relative stiffness differences between adjacent zones, these examples illustrate how deformation may be proactively constrained to particular regions, thereby minimizing undesired deformation in higher stiffness areas. A geometric net refers to a 2D object which may be modified in order to form a 3D shape [46]. Figure 5A shows a set of triangular geometric folding nets, while Figure 5B shows a set of folding box nets in both undeformed and deformed states. For the deformed states shown in Figure 5C and 5D these states were achieved by affixing the appropriate free edges of the nets as seen in Figure 5A and 5B using fishing wire. No other external influences besides the edge-to-edge fixture and the stiffness zones were applied to alter the final state of these deformed nets. The curvature observable in both specimens with and without hinges is entirely due to boundary conditions applied over the flexible VTP material. The difference in stiffness within the included joints vs the adjacent material allows for a significant reduction in deformation of the higher stiffness faces, thus resulting in a greater agreement between intended geometry and final geometry based on the undeformed net arrangement.

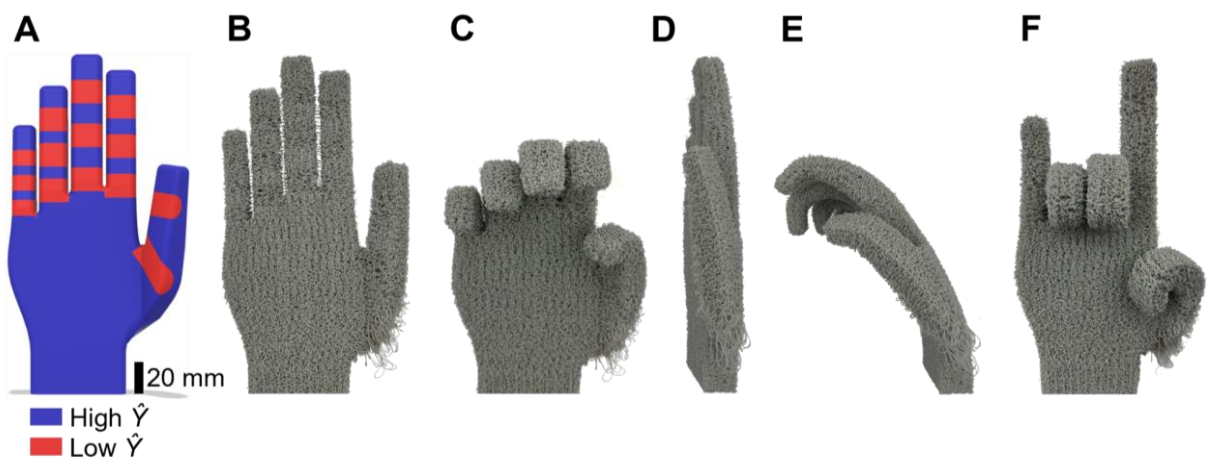

**Figure 6.** Example of VTP hand cable actuated with fishing wire threaded along the surface of the palm and fingers, which utilizes $\hat{Y}$ zones placed along typical joint locations for selective deformation A) 3D model used to manufacture printed hand color coded according to assigned $\hat{Y}$ zones. B) Front view of undeformed hand. C) Front view of deformed hand. D) Side view of undeformed hand. E) Side view of partially deformed hand. F) Front view of selectively deformed hand in the form of a complex gesture.

Programmed VTP material properties can be applied volumetrically, enabling programmed 3D deformation, as seen in Figure 6. Much like in thin 2.5D materials, volumetric regions of low stiffness preferentially deform and bend when adjacent to regions of higher stiffness material. By controlling the location and degree of this stiffness differential, VTP produces programmable deformation and bending within arbitrary geometry and conditions. As a demonstration, we created a foam approximation of a human hand as seen in Figure 6. The joints of the hand were achieved by lowering the stiffness of semi-cylindrical zones at the location of each joint found in a typical hand (Figure 6 A). A simple cable driven system capable of emulating a variety of gestures as seen in Figure 6 B-F was then created by threading fishing wire up through the already porous surface of the palm and fingers. This system was able to individually contract each finger in a simple downward motion. Additional degrees of freedom may be enabled by adding additional cables along the external surface of desired motion.

## 2.5. Common Foam Emulation

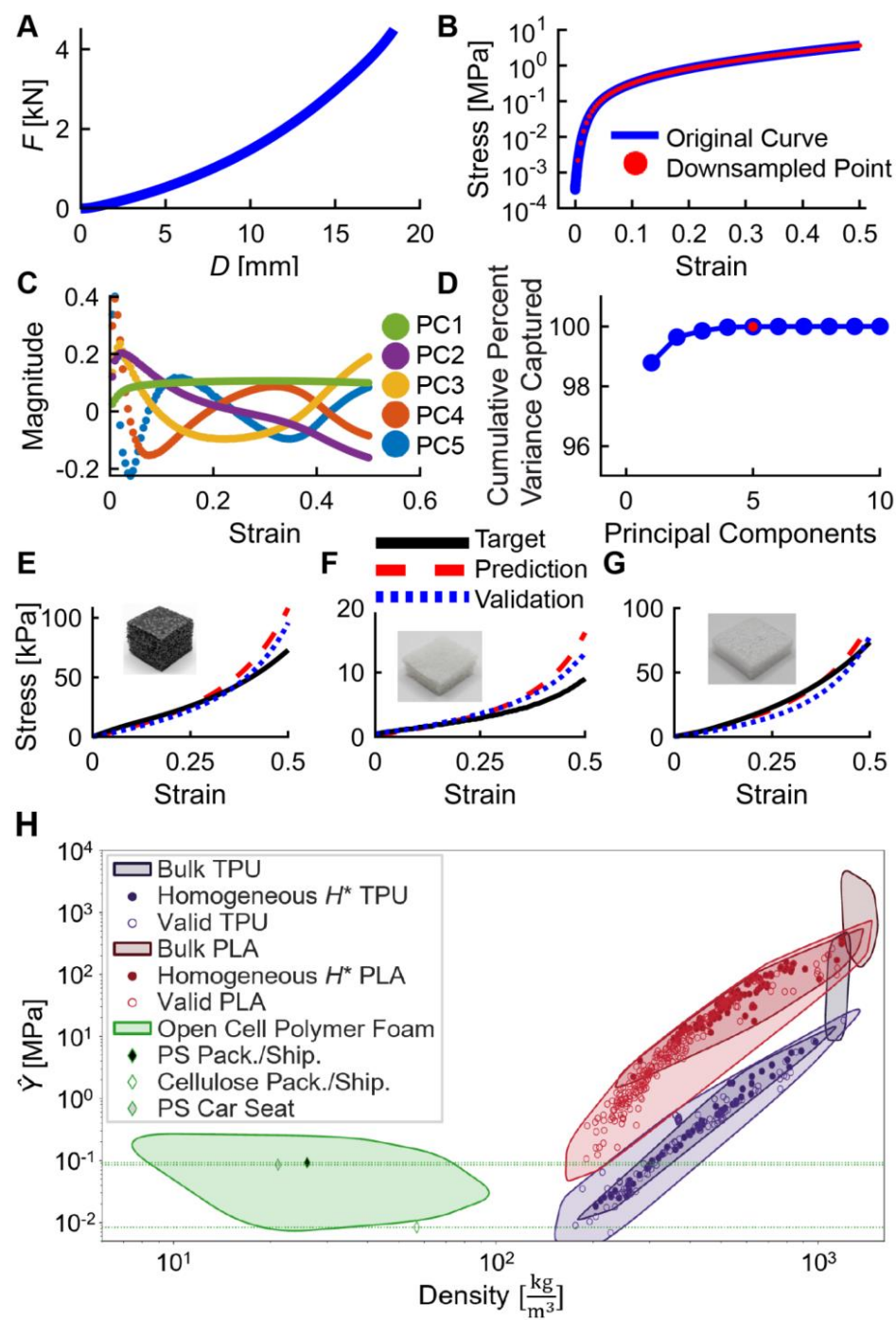

**Figure 7.** A) Force ($F$)-displacement ($D$) of a TPU cube compressed to a 4.5 kN force limit. B) Log stress-strain curve (blue) converted from F-D curve is down sampled to 100 points (red) equally

spaced to 50% strain. C) Principal component analysis breaks down sampled curves into 100 components (only top five shown for clarity). D) Cumulative percent variance captured by number of components. The vast majority of variance is captured by the first five components. E-G) Stress-Strain curves for three foam samples (black curves), the sample predicted to most closely match their performance (red dashed), and the tested performance of that predicted sample (blue dotted). H) Ashby plot containing all relevant density and modulus data taken by this study for PLA and TPU, with additional points for reference foams and envelopes for bulk PLA, TPU, and open cell polymer foams as defined by level 3 Ansys materials dataset.

In addition to controlled deformations, we hypothesized that the processing freedom afforded by VTP would allow VTP foams to replicate the material performance of common foams. To explore this, we ran an additional campaign with the SDL to explore VTP print parameters for TPU filament. As before, the campaign used active learning to fully explore the parameter space with 155 experiments. Next, F-D curves (Figure 7A) were converted to stress-strain curves based on the sample geometry. The stress-strain curves were then down sampled to 100 stress points that were equally spaced in strain from $\varepsilon = 0.005$ to 0.5 (Figure 7B). We then took the logarithm of these stress values to preclude the possibility of predicting negative stress. Principal component analysis was then applied on all 155 TPU samples tested (Figure 7C-D). A separate GPR model was conditioned on the scores for each of the first five principal components as a function of their processing conditions. These GPR models could then be used to predict the performance of VTP foams that have not yet been tested.

In order to evaluate the VTP foam selection pipeline we selected three reference foam samples from everyday objects, consisting of a polystyrene (PS) foam used for packing/shipping (Figure 7E), a cellulose foam used for packing/shipping (Figure 7F), and a PS foam used as part of a children's car seat (Figure 7G). They were tested in quasi-static compression using the same protocol as the VTP foams. The performance of each foam was then compared to the predictions of 125,000 potential VTP experiments selected using Latin hypercube sampling. An error function minimized the mean squared error of the difference between the log stress prediction and the foam stress. The VTP cube with the lowest error was selected and tested, showing the ability to match the performance of common foam samples.

VTP manipulates microstructure and its dependent mechanical properties by introducing void space. Therefore, the bounds of the mechanical property space are directly related to the bounds of the porosity of the foam. As there is no theoretical maximum $H^*$, and consequently no theoretical maximum cell size, the only limit to how much the density of the material can be reduced from bulk is $D_t$ as the minimum $H_L$ and length scale separation of cells and the overall part. In order to fully visualize the expanded range of stiffnesses which VTP enables, as tested on the SDL system, a comprehensive Ashby plot was created for estimated modulus vs density as seen in Figure 7H. This plot reveals that, within the limited scope of the data collected, the range of predicted moduli was extended downward from bulk by approximately four orders of magnitude for both PLA and TPU. From the perspective of modulus, the range of moduli enabled by VTP effectively spans the bulk modulus of each material and the conventional modulus space occupied by open cell polymer foams as defined by the Ansys level 3 materials dataset. Furthermore, it is hypothesized that these regions do not represent the maximum achievable range of mechanical properties for these materials, as the data collected did not include all possible VTP parameter sets. Thus, further research may reveal a wider range of microstructures and their associated mechanical properties.

## 3. Conclusions

This work demonstrates a transformative way of spatially varying the material properties of foams using simple desktop printers. By harnessing viscous thread instability to enable VTP, we enable localized customization of mechanical properties by creating gradient microstructures. These gradients demonstrate spatially variable capabilities beyond what is currently achievable via conventional chemical foam manufacturing. Due to the correlation between microstructure and mechanical properties in cellular materials, the ability to spatially vary microstructure results in spatially variant mechanical properties, including localized stiffness, porosity, and deformation in three dimensions. This VTP process is enabled in a significantly expanded property space by deploying a predictive model that explores a newly discovered homogeneous subspace, linking input VTP parameters to output foam material properties without compromising structural intent. Additionally the exploration of this subspace revealed an underlying self-stabilizing characteristic of VTP layer thickness, allowing for an acceptable margin of error when selecting VTP characteristics that affect overall layer geometry. This model was produced using a SDL which combined automated experimentation with machine learning. The application of this model enabled us to replicate the stiffness of a selection of foams used in packaging, padding, and cushioning, suggesting that VTP may be used to emulate significant characteristics of commonly distributed conventional foams. The predictive mapping models were developed and validated for TPU and PLA, suggesting these models may be trained on any material compatible with ME 3D printing. Overall, these findings indicate that VTP offers a versatile and precise method for manufacturing foams with tailored properties beyond the current state of the art, potentially paving the way for a wide variety of new application spaces for foams and 3D printed materials.

## 4. Experimental
### 4.1. G-code Generation

G-code files for 3D printing of test specimens and demonstration samples were generated using two different methods. For the test specimens, a Python program generated G-code for a cube with side length $L$ and hardcoded a rectilinear infill. Parameters $V^*$, $H^*$, $\Delta L$, $\Delta Z$, and $\alpha$ were input and applied homogeneously, resulting in a uniform, cubic test specimen. The second method involved a custom slicer capable of processing an arbitrary number of non-overlapping STL mesh files with varying geometries. This was used for all samples with inhomogeneous densities. Each imported mesh was assigned a local $V^*/H^*$ pair, while a global $\Delta L$, $\Delta Z$, and $\alpha$ were prescribed for the entire print. The slicer generated GCode with a rectilinear infill for the union of all meshes, refined the toolpath, and assigned relevant G-code values ($Z$, $E$, and $F$) based on the local $V^*/H^*$. This resulted in a single toolpath for arbitrary geometry with GCode parameters specified by the local $V^*/H^*$ values.

### 4.2. Layer height measurements

Layer height measurements (Figure 2F) were made through a manual process. First, a single rectangular prism of $H = 60$ and $L = 30$ was printed from PLA for $H^* = 3$, $V^* = 0.3$, $\Delta L = 1.5$ and $\Delta Z = 2.2$. The height of the cube was measured using a universal testing machine (UTM) (Instron 5965) by lowering the platen manually until the force reached 1 N, ensuring that stray strands of filament did not artificially increase the height measurement. The height of the cube was taken as the platen separation of the UTM. Then, the bottom layer of the cube was removed with wire cutters. Because the filament is more tightly bonded within the layer than between layers, it was possible to cut away a single layer without disturbing the remaining layers. However, because of the thin nature of the removed layer, it was destroyed in the process. The height of the remaining cube was then measured in the UTM with the 1 N threshold. By comparing the difference in height between the two measurements, the height of the removed layer was inferred. This process was repeated until only two layers remained. Unfortunately, because the final layers were so thin, measurement of the last two layers was not possible. This process was repeated for the other two cubes with $H^* = 6$ and 9.

### 4.3. Automated testing

Automated testing leveraged the Bayesian experimental autonomous researcher (BEAR) previously developed [34,35]. The system consists of a UTM (Instron 5965), a scale (Sartorius CP225D), a material extrusion (ME) printer (Prusa Mk3S+), and a six-axis robot arm (Universal Robotics UR5e).

The learning loop consisted of several steps performed in sequence. First, an experiment is selected using active learning. GPR models were conditioned in MATLAB using the built-in function fitrgp using a squared exponential kernel with automatic relevance detection. Potential experiments were calculated using Latin hypercube sampling and an experiment was selected using a maximum variance decision-making policy. Initially, $Y$ was used as the target metric for the GPR, but it was later changed to $H_L$ to ensure that both metrics were mapped fully.

G-code for the selected experiment was then generated using a custom python script. The g-code was sent to the printer using OctoPrint through the python package OctoRest. For PLA (eSun PLA+ gray), printing temperature was set to 215 °C and the bed temperature was set to 60°C. For TPU (Ninjatek Ninjaflex blue), printing temperature was set to 230 °C and the bed temperature was set to 50 °C. For both materials, the part was removed by the robot after the bed had cooled below 32 °C.

The part was then transferred to the scale where it was weighed. If the measured mass was more than 5% from the expected value, the part was discarded due to the likelihood of a print error. After recording the mass, the part was transferred to the UTM for testing. During testing, the UTM lowered its platens at 2 mm/min until the force reached 4.5 kN or the platens were separated by less than 0.4 mm.

The modulus was calculated by converting the force-displacement curve into a stress-strain curve. The height for this conversion was measured by the UTM as when the force exceeds 0.3 N. The cross-sectional area was assumed to match the designed dimensions for the cube of 30 x 30 mm$^2$. The modulus was then calculated using a linear fit of the stress between 5% and 15% strain to avoid any toe regions.

The $H_L$ of the part was calculated by taking the height of the final part and dividing by the number of layers. For a part to be considered homogenous, $H_L$ must be within 5% of $\Delta Z$.

### 4.4. Principal component analysis

The stress-strain curve was converted to log10 and truncated at 50% strain. It was then down sampled to 100 equally spaced points. PCA was performed using MATLAB's built in pca function. Because 99.99% of cumulative variance was captured by the first five components, only the first five components were used for stress-strain curve prediction. For each of these components, a GPR model was conditioned to predict the score of the component based on the four-dimensional input space ($V^*$, $H^*$, $\Delta L$, and $\Delta Z$). Sampling points (125,000) were then selected using Latin hypercube sampling for $V^*$, $H^*$, and $\Delta L$. $\Delta Z$ was selected using the $H_L$ GPR model to set $\Delta Z = H_L$, therefore ensuring that the resulting sample was homogenous. The predicted score components were then transformed back into the log10 stress/strain space. The predicted curves were compared to the foam samples using the error function shown in equation 3. The closest curve was selected for each foam type and tested.

$$Error = \frac{\sum (log(\hat{\sigma}_n) - log(\sigma_n))^2}{n} \qquad (3)$$

Foam samples were cut to 30 x 30 mm$^2$ rectangular prisms to match the cross-section of the sample VTP foams, but the heights of the foam samples were not modified. The foams were weighed by hand and manually transferred to the UTM. The UTM tested them using the same 2 mm/min speed and the same 4.5 kN stop threshold protocol.

**4.6. Local Effective Stiffness Testing**

For the orthotic stiffness heatmap as seen in Figure 4D. A local effective Young's Modulus method was employed over 199 data points, evenly distributed 20 mm apart in both X and Y directions across the surface of the orthotic. This data was collected utilizing an Instron 68SC-2 Universal Testing System equipped with a 2 kN load cell affixed with a flat circular 10 mm diameter probe printed on a Carbon M1 out of UMA-90 [23], mounted axially above a 10 kN testing anvil. The orthotic was printed on a standard Prusa Mk4 using NinjaTek NinjaFlex TPU. The resulting orthotic was approximately 6 mm in thickness.

The procedure for testing included an initial 1 mm/s ramp until a force threshold of 1 N was met to establish height of initial contact. The probe then returned 1 mm to pause for one half second before starting data collection at 6 mm/min for a total of 2 mm deformation.

The local effective modulus was calculated by converting the force-displacement curve into a stress-strain curve. The term "local effective modulus" refers to the fact that this testing method used a probe that did not fully envelope the cross section of the orthotic to sample the average mechanical stiffness of a small localized region of the overall part. The cross-sectional area was assumed to match the designed dimensions of the probe of $\pi$ x $5^2$ mm$^2$. Young's modulus was calculated using a linear fit of the stress-strain data between 10% and 15% strain to avoid any toe regions. Due to the fact that the material being tested in this method is not isolated from the surrounding material, it is expected that the resulting data will be influenced by boundary effects.

**Acknowledgements**

This work was funded by: a gift from the Ford Motor Company, NSF Grant number 2212049, the National Defense Science and Engineering Graduate Fellowship (NDSEG) program. KAB and KLS acknowledge support from the National Science Foundation (DMR-2323728) and